# LOCALIZED SURFACE PLASMON RESONANCE (LSPR) AND REFRACTIVE INDEX SENSITIVITY OF VACUUM EVAPORATED NANOSTRUCTURED GOLD THIN FILMS

V. N. Rai[1], A. K. Srivastava[1], C. Mukherjee[2] and S. K. Deb[1]

[1]Indus Synchrotrons Utilization Division

Raja Ramanna Centre for Advanced Technology

Indore-452013, (India)

[2]Mechanical and Optical Support Section

Raja Ramanna Centre for Advanced Technology

Indore-452013, (India)

**Email: vnrai@rrcat.gov.in**

**Phone:+91-731-2488142**

**FAX:  +91-731-2442140**

**ABSTRACT**

The plasmonic properties of vacuum evaporated nanostructured gold thin films having different types of nanoparticles are presented. The films with ≥ 6 nm thickness show presence of nanorods having non cylindrical shape with triangular base. Two characteristics plasmon bands have been recoreded in absorption spectra. First one occurs below 500 nm and other one at higher wavelength side. Both the peaks show dependence on the dielectric property of surroundings. The higher wavelength localized surface plasmon resonance (LSPR) peak shifts to higher wavelength with an increase in the nanoparticle size, surface roughness and refractive index of the surrounding (Methylene Blue dye coating). This shows that such thin films can be used as sensor for organic molecules with a refractive index sensitivity ranging from 250 - 305 nm/RIU (Refractive Index Unit).

## 1. Introduction

The plasmonic properties of nanoparticles have attracted much attention in the past decade [1-5]. Their potential use in optical spectroscopy, photonic devices, biosensors, as well as in the waveguides makes them a very active area of research. The localized surface plasmon resonance (LSPR) of such nanoparticles plays an important role in such applications. Localized surface plasmon resonance occurs when the electrons in the nanoparticles interact with electromagnetic radiation. The noble metal nanoparticles such as gold (Au) and silver (Ag) provide strong extinction and scattering spectra due to this property that has been utilized for various important applications in different field of research, particularly for surface enhanced spectroscopy and sensors [3, 6-12]. These plasmonic materials have been proved to be a promising candidate for the design of structures with extra ordinary optical properties [13-14]. Various types of nanostructures have been used to provide LSPR such as nanoparticles, nanorods, nanoporous films as well as array of nanoholes [15-20]. All these structures show LSPR peak at different locations. Because LSPR frequency depends mainly upon the nanoparticle size, shape, material properties and surrounding medium [16-17, 21-24]. Particularly nanorods show two resonance peaks corresponding to the transverse and the longitudinal resonance mode. Transverse mode resonance occurs at fixed wavelength (~520 nm), whereas longitudinal mode wavelength shifts towards near infrared with an increase in its size and refractive index of the surrounding medium [25]. Two peaks have also been observed in the case of nanoporus gold thin film, first one due to interband transition and the other one due to LSPR. The position of interband transition (at lower than transverse mode wavelength ~520 nm) has been found at different locations by different groups [17-18]. However no information is available about the non-cylindrical shape of nanorods. Mostly cylindrical nanorods have been obtained by chemical methods and have been studied in liquid medium. Very less information is available about gold nanorods presents in thin films, where particles are present in sufficiently high density such that their electric fields overlap. Normally optical properties of the thin films depend mainly on the process of its preparation [26-28]. The nanostructures of the vapor deposited films are supposed to be columnar. The packing density of the column is dependent upon the energetics of the condensation processes and consequently upon the deposition methods. Vacuum evaporation method has been used eralier to obtain the nanostructured gold thin films and has been reported that mainly the size of the nanoparticles in x-y plane decides the optical properties (LSPR) of the thin films [28]. Link et al. [25] have derived a relation between aspect ratio of the nanorods and its LSPR peak wavelength. However it is not clear how these relations behave in the case of non-cylindrical nanorods. Thus better understanding of plasmonic properties of non-cylindrical nanorods in thin film is essential for producing optimized devices for surface enhanced spectroscopy and plasmonic sensors.

Measurement of shift in localized surface plasmon resonance (LSPR) absorption peak in thin films allows study of changes in the optical properties of the nanostructured materials. Nanoparticles assembled in the form of adsorbed layers particularly on the thin films have coupling in their electric fields that influences the localized plasmon resonance frequency of the films. In such a situation changes in the LSPR peaks of the composite films are correlated with changes in the dielectric properties ($\varepsilon$) of the materials, that is, the change in its refractive index (n) and extinction coefficient (k). This property has been exploited as plasmonic sensors. The cylindrical nanorods show linear variation in LSPR peak with change in the refractive index (RI) of the surrounding medium. The refractive index sensitivity of the cylindrical gold nanorods of aspect ratios 2.5-4.2 changes from 216 to 352 nm/RIU (refractive index units) [15]. However it is important to compare the refractive index sensitivity of the non-cylindrical nanorods with other kinds of nanostructures.

The main aim of this paper is to study the optical and structural properties of vacuum evaporated nanostructured gold thin films, where nanoislands are in the form of non-cylindrical nanorods. Refractive index sensitivity of the composite film having coating of Methylene Blue dye has been obtained and compared it with the RI sensitivities of other nanostructures obtained by different methods. Effect of surface roughness of the thin films on its optical properties (LSPR) is also discussed.

## 2. Experimental

The preparation of nanostructured thin films is critical for this type of experiment because optical properties of the films are very much dependent on the size, shape and particle density of nanoislands in the films. The experimental conditions such as rate of film deposition, type of substrate, the substrate temperature and the required metal film thickness play an important role in deciding the properties of films. For this experiment gold nanostructured films are deposited on the glass substrate using vacuum evaporation technique. The vacuum deposition system is evacuated at chamber pressure of $\sim 3 \times 10^{-5}$ mbar using a diffusion pump. A resistively heated tungsten boat is used for evaporating the gold for thin film deposition. The thickness and the rate of deposition of films are monitored by using a quartz crystal microbalance. The glass substrates are thoroughly cleaned before the film deposition in order to get exact and reproducible information about the location of LSPR in the broad spectral range. The substrate temperature, deposition rate and the film thickness are the important parameters, which are controlled during the film deposition. In the present experiment deposition parameters are chosen as $25^0$ C, $< 0.01$ $A^0 s^{-1}$ and 5-150 $A^0$ respectively. The most important parameter as deposition rate ($\sim 0.01$ $A^0$ $s^{-1}$) is kept very slow in order to grow individual/separated gold nanoislands. The main idea was to produce nanoislands in the form of nanorods, because with increase in thickness the height of nanoisland increases. For each deposition, tungsten boat is slowly heated such that vaporized gold got deposited on the glass substrate before melting of the gold. After film deposition, system is allowed to cool gradually near to room temperature. Testing results shows that once the substrate temperature and deposition rate are kept constant, the deposited films provide similar characteristics. Some of these films are coated with a thin layer of Methylene Blue dye by dipping the gold film in an ethanolic solution of dye. Normally thickness of the dye layer produced by this technique varies linearly with the concentration of the dye solution.

The optical absorption spectra of the bare gold thin films and its composite films with dyes are recorded in transmission mode using UV-Vis spectrophotometer (Lambda 20 Perkin Elmer) in order to find information about the LSPR in both types of films. The spectra are recorded in the wavelength range from 190 – 1100 nm in the steps of 0.3 nm and a resolution of 0.1 nm. The shape and size of the nanoislands in thin film of gold are determined by transmission electron microscopy (TEM) and atomic force microscopy (AFM). TEM measurements are carried out on PHILIPS CM-200 microscope equipped with $LaB_6$ filament. The line resolution of the microscope is 1.4 Å. Surface morphology of thin films are measured using an atomic force microscope (NT-MDT, SOLVER- PRO) having Si cantilever tips (resonant frequency-190 KHz, spring constant ~ 5.5 N/m) in non-contact mode. Radius of curvature of tip is ~ 20 nm.

### 3. Results and Discussion

### A. LSPR in Gold Thin Films and its Composite Film with Methylene Blue

The optical absorption properties of gold nanoparticles in the visible range are represented by the effect of boundary conditions of the coherent electron oscillations and by the interband d→sp electron transition. Fig.-1 shows the absorption spectra of the nanoparticles present in the thin film of gold having thickness 0.5 to 15 nm. These spectra are recorded in the wavelength range from 190- 1100 nm. Each film shows two resonance absorption peaks, where shorter wavelength peak is defined as $\lambda_1$ and the longer wavelength one as $\lambda_2$. Both the peaks shift to higher wavelength side with an increase in the thickness of the films as shown in Fig.-2. Particularly $\lambda_1$ peak shifts slowly at lower thickness of films, whereas it shows smaller shift or saturation towards higher thickness films. The $\lambda_1$ peaks are smaller in amplitude, whereas $\lambda_2$ peaks are broad and have higher amplitude. Normally spherical nanoparticles show only one resonance peak due to LSPR, which shifts to higher wavelength side and broadens with an increase in its size. However two resonance peaks have been observed in the case of nanorods where lower wavelength peak occur due to transverse mode and higher wavelength one due to longitudinal mode. The peak due to transverse mode remains nearly at fixed locations ~520 nm due to its cylindrical shape, which is coincident with the plasmon band of spherical particles. The longitudinal mode peak shifts and broadens to higher wave length side with an increase in the aspect ratio of the nanorods [25]. Even nanoporous gold films also shows two peaks where lower wavelength peak is assigned due to interband transition, which occurs below 500 nm. Normally it also remains fixed. But different groups have reported its location varying between 400-500 nm [17-18]. In this case higher wavelength peak is reported to be due to LSPR. In the present experiment lower wavelength peak $\lambda_1$ has been observed below 400 nm. It shifts to higher wavelength side as the thickness of the films increase upto 6 nm and then remains nearly same. Here second peak $\lambda_2$ is clearly assigned due to LSPR. The presence of two peaks indicates that nanoparticles in the present thin film may be having nanorods like structures. However variation in the location of $\lambda_1$ peak with film thickness indicates that the nanorods present in the thin films may not be having cylindrical shape. It has been reported that LSPR occurs due to dipole resonance, whereas multipole resonance can create peaks at different location depending on the size and shape of the nanoparticles [22-25]. The shift and broadening in $\lambda_2$ is due to change in the size and shape of the nanoparticles probably the nanorods. This indicates that measurement of exact size and shape of the nanorods present in the thin films is important in order to better understand the correlation between optical and structural properties. Next section discusses the size and shape of the nanoparticles present in these films.

In order to see the effect of dye coating on the optical properties of gold thin films particularly on the LSPR peak ($\lambda_2$), absorption spectra of Methylene Blue dye coated on gold thin films of thickness 3, 6 and 10 nm have been recorded and is shown in Fig.-3. The concentration of Methylene Blue is kept low as $10^{-4}$ M/l. Absorption spectra show that LSPR peaks $\lambda_2$ of the gold thin films split into two peaks in the case of dye coated composite films, which is as expected [29-30]. Here the first peak is due to dye absorption whereas the second one is due to LSPR. The splitted peaks are found at different locations, than 635, 677 and 736 nm observed for 3, 6 and 10 nm gold film respectively or 668 nm for Methylene Blue dye. The splitted peaks in composite films of 3, 6 and 10 nm occur at 595 and 668 nm, 599 and 733 nm as well as on 608 and 791 nm respectively. Here 3 nm composite film shows a second split peak at 668 nm, which seems to be due to dominant absorption by dye in the case of very thin gold film. In the case of composite films of 6 and 10 nm thickness splitted peaks are observed at quite different wavelength than LSPR and dye absorption peaks. Broadening in the peaks becomes dominant over the splitting in the case of composite films of higher gold thickness. This is because of broad LSPR peak in the case of thick gold film.

Variation in LSPR peak with an increase in gold film thickness in the case of gold thin film and its composite with Methylene Blue dye is shown in Fig. - 4. In both the cases LSPR peak increases linearly with the gold film thickness. There is a clear shift in the LSPR peak in the case of dye coated composite films than bare gold films. This is due to change in the dielectric properties of the thin films of gold after dye coating.

Particularly this variation is due to change in the refractive index of the surrounding medium around the gold nanoparticles, which changes the resonance condition and the location of LSPR peaks. The effect of refractive index on LSPR peaks will be discussed in detail in coming sections.

B. **Surface Morphology of the Gold Thin Films**

It has been discussed above that the LSPR wavelength not only depends on the film thickness but on the other deposition parameters, which decides the shape and size of the nanoislands in the films. The morphological features of the thin films also depend on the film thickness as well as on the other parameters such as deposition rate, temperature and atmosphere during film deposition as well as on the annealing process after the film deposition. Transmission electron microscopy (TEM) has been used to find the size and shape of the nanoparticles present in the gold thin film obtained by vacuum evaporation technique. The TEM images of the gold films as well as distribution of ellipticity of nanoparticles for 1, 3 and 6 nm thick films are shown in Fig.-5-7. Here ellipticity has been defined by the ratio of major to minor axis of the nanoparticles. It shows that 1 nm thick film has mostly nanoparticles with ellipticity one. In this case most of the nanoparticles may be having spherical shape. As the film thickness increases to 3 nm, ellipticity of the nanoparticles start changing but still more number of particles are having ellipticity one (Fig.-6). It seems this is the point from where sphericity of the nanoparticles starts changing. The 6 nm thick film (Fig.-7) contains nanoparticles of different shapes like sphere, hexagonal, triangle, rhomboid, tadpoles, nanoboots, nanorods and turmeric shapes with maximum number of particles having ellipticity 1.4. The radius distribution shows that 1 nm thick film has maximum numbers of particles having radius from 1.75 to 2.75 nm, whereas it is 4-5 and 5-7 nm for 3 and 6 nm thick films respectively. This indicates that the film with less thickness (< 6 nm) has nanoparticles of nearly spherical shape, which got distorted as the film thickness increases. In fact, the average ellipticity increases with an increase in the film thickness. The film of thickness ~6 nm has average ellipticity of ~1.7 as shown in Fig.-7b. AFM images of these films are recorded to find the exact shape of the nanoparticles particularly in three dimensions in other words morphology of the film surface. Fig. - 8 shows 2 and 3

D AFM pictures of the 3 nm thick film. No clear base structure is seen in two dimensional AFM pictures of the films. However 3D picture shows an increase in the height of nanoparticles. Fig.-9 shows 2 and 3 dimensional AFM images of the ~6 nm gold film. In this case 2 D pictures clearly shows the presence of triangular shaped nanoparticles, whereas 3 D picture shows clear vertical structure of the nanoparticles in the form of nanorods. This shows that sufficient numbers of nanorods are present in ~6 nm thick films, but has non cylindrical shape. Formation of nanorods becomes dominant for the film thickness of > 3 nm. The growth of nanorods seems to be possible due to slow deposition rate, where evaporated material got stacked one over the other without any diffusion. This indicates that the nanoparticles present in the film of different thicknesses may be having nanorods of different aspect ratio with a certain possible distribution. This is in agreement with the earlier reported results that formation of nanorod structure during vapor deposition process is possible after crossing the percolation transition [31]. The coverage of the surface or particle density increases with an increase in the film thickness (Fig. 5 to 7), which is measured in the terms of volume fraction and is found ranging from 0.25 to 0.53 for films of thickness 1 to 10 nm respectively. The roughness parameters of ~3 and ~ 6 nm films are also measured using AFM and is shown in Table -1. This shows that average height, average and rms roughness of the films increase with an increase in the film thickness from ~3 to ~6 nm. A comparison of these data with that of LSPR peak wavelength shows that peak shifts to longer wavelength with an increase in the island size, volume fraction and surface roughness of the thin film. These results are in agreement with the earlier reported results, where different island size and roughness in the films were created by annealing the thick films [27]. An increase in the LSPR peak wavelength with an increase in the deposition rate also seems to be due to an increase in the surface roughness in the films. The above results further indicate that all the parameters of film preparation play an important role in deciding the surface roughness and /or size of the islands of the thin film and consequently its LSPR wavelength.

As it has been observed above that the nanorods present in the films are in the form of non-cylindrical shape with an elliptical (triangular) base. Therefore, aspect ratio of the nanoisland particles has been defined here as the ratio of the film thickness and the average ellipticity of the nanoparticles in that film. Here it is important to find out a correlation between aspect ratio of this type of nanorods with its LSPR peaks. An expression has been reported by S. Link et al. [25] having correlation between aspect ratio (R), LSPR peak ($\lambda_m$) and the dielectric constant of the surrounding medium ($\varepsilon_m$) as

$$\lambda_m = (33.34R - 46.31)\,\varepsilon_m + 472.31 \qquad \text{------------} \quad (1)$$

This relation was derived considering that the real part of the gold dielectric function is decreasing nearly linearly with the wavelength of the light in the range between 500 and 800 nm and also that $\lambda_m$ changes linearly with R. They found that to fit the experimental data to eq. (1), $\varepsilon_m$ should be size dependent and increasing in a nonlinear fashion with decreasing the aspect ratio of the nanorods. We have used the data of present experiment such as aspect ratio and $\lambda_m$ in eq. (1) and calculated the value of $\varepsilon_m$ for each film as shown in table-2, which shows variation with film thickness as expected. The value of $\varepsilon_m$ obtained by this method for non-cylindrical nanorods is in close agreement with the earlier reported results for cylindrical gold nanrods [16]. Fig. - 10 shows the variation of $\lambda_m$ with aspect ratio of the nanorods, which is plotted for different value of $\varepsilon_m$ ranging from 1 to 16 as per eq. (1). Experimentally obtained data of aspect ratio and $\lambda_m$ for non-cylindrical nanorods are also plotted in the same figure. Experimental as well as theoretical data show a linear variation with a difference that there is a mismatch towards lower aspect ratio side. This mismatch may be due to negative value of dielectric constant for lower values of aspect ratio (<2) as shown in table-2 and Fig.-10. This indicates that above expression is not fitting the data for nearly spherical nanoparticles with aspect ratio close to one. In contrast to linear variation of LSPR peak with aspect ratio, when ellipticity and the volume fraction are used in place of aspect ratio it provided slight deviation from linearity near ~10 nm film thickness. It seems that decrease in inter particle distance with an increase in film thickness increases the particle-particle interaction that affects the LSPR redshift as well as broadening even more than linear, which may be the reason behind nonlinearity with ellipticity and volume fraction.. This also may be due to multipole oscillations in nanoislands with an increase in the thickness of the films. This observation is in agreement with the earlier reported results that the LSPR peak wavelengths for the gold nanoisland films are mainly dependent on the two dimensional size of particles in X-Y plane [28]. However, the above results indicates that both the aspect ratio as well as the particle size of nanoislands in XY plane play an important role in deciding the LSPR peak wavelength in nanorods.

## C. Refractive Index Sensitivity of Composite Gold Film with Methylene Blue Dye

The optical property of the thin film is mostly decided by dielectric constant of the surrounding medium ($\varepsilon_m$) as discussed in previous section, which consists of two important parameters as coefficient of refraction (n) and the extinction coefficient (k). In the case of thin films having nanoislands, the effective dielectric constant can be defined in the terms of volume fraction and the polarization of the metallic nanoparticles [32-33]. Polarization of the film is proportional to the number of particle times volume of each nanoparticle, that is, fractional volume (F) of the film. It has been proved to be more important parameter in expressing the optical properties of the film as distribution of the nanoparticles cover different area in the films. Based on the free electron model of the metal, it has been reported that the localized surface plasmon resonance (LSPR) absorption maximum ($\lambda_m$) and the wavelength of the plasmon resonance peak of bulk metal $\lambda_p$ (for gold 131 nm) is related with the volume fraction (F) as well as the refractive index ($n_s$) of the effective medium surrounding the films and is given by [33]

$$\lambda_m = \lambda_p \left[ 1 + \left( \frac{2+F}{1-F} \right) n_s^2 \right]^{\frac{1}{2}} \qquad \text{-------------------} \quad (2)$$

This is a simple analytical approximation for the surface plasmon band shift within the frame work of Maxwell Garnet theory [32]. Variation of $\lambda_m$ with volume fraction (F) for different value of refractive index ($n_s$) ranging from 1 to 3 based on eq. (2) is shown in Fig.-11. It shows that $\lambda_m$ first increases slowly with an increase in volume fraction (film thickness). This also shows that LSPR shifts to higher wavelength with an increase in refractive index. The experimentally observed values of $\lambda_1$ and $\lambda_2$ for different volume fraction (film thickness) are also plotted in Fig.-11. This shows that LSPR peaks at $\lambda_2$ best fit the theoretically calculated plot for $n_s$ close to 2.4, whereas $\lambda_1$ peaks best fit for $n_s$ between 1 and 1.5. The refractive index 1 and 1.5 corresponds for air and glass respectively. The complex refractive index of gold thin film has been reported as ~2.5, which decreases with film thickness [34]. It shows that some of the nanoparticles in the thin films feel surrounding refractive index dominated by air and glass. It seems to be possible in the case of smaller nanoparticles, where inter particle separation is more and the medium between particles is glass and air. However bigger nanoparticles may be showing effective refractive index ~ 2.4, which is close to the complex refractive index of gold thin films [34]. This may be possible due to bonding between gold nanoparticles with different air molecules. In the case of less inter-particle separation particle –particle interaction also increases. This indicates that $\lambda_1$ peaks observed in the case of gold thin film seems to be due to resonance of the smaller nanoparticles having larger inter-particle separation filled with glass and air around them. Such resonance peaks at lower wavelength is also possible due to multipole resonance in the particles having different shapes (Many edge) as has been reported earlier [17, 24]. So, there may be a contribution from transverse mode resonance of nanorods, which seems to be dominated by multipole resonance of different shapes present in the base of the nanorods. However further investigation is needed in this direction to clarify these points.

It is well known that coupling of other molecule with the nanoparticles change the LSPR wavelength of the nanoparticles. In this case Methylene Blue dye has been coated on the gold thin films, which shows clear shift in LSPR resonance to higher wavelength side. Fig.-12 shows that experimental data points of LSPR of composite films also fit the curve obtained using eq.-(2) for refractive index close to 2.5. Table- 3 shows the values of $\lambda_m$ in the case of 3, 6 and 10 nm gold and Methylene Blue coated composite gold thin films. The values of refractive index calculated using eq. - 2 are also shown for gold as well as composite film of gold and dye in table-3. It shows that values of refractive index of surrounding increases in the case of dye coating on gold nanoaprticles. This clearly indicates that such thin films of gold can be used as a refractive index sensor. Refractive index sensitivity ($d\lambda/dn$) of these films is calculated and plotted against film thickness and is shown in Fig.-13. This shows that sensitivity of such films increases from 250 – 305 nm/RIU (Refractive Index Unit) as the film thickness increases from 3 to 10 nm. Sensitivity shows saturation effect towards higher thickness of gold thin film. Saturation may be occurring due to increase in particle-particle interaction in the case of thick films. Similar refractive index sensitivities have been reported in the case of cylindrical nanorods of gold as well as in the case of nanoporous gold film [15, 18]. This shows that non cylindrical nanorods present in thin films of gold and obtained by vacuum evaporation technique can also be used as refractive index sensor in the visible range with equally good sensitivity. This technique for preparation of nanostructured thin gold films is comparatively easy and can be beneficial.

Finally this study indicates that it is possible to find a thin film of predefined optical properties (LSPR peak wavelength) using vacuum evaporation technique with predefined deposition parameters. However, further investigation is needed to find different surface structures and its preparation techniques in order to increase the sensitivity of films in visible spectral range for its better application as a sensor.

## 4. Conclusions

In summary, we can say that thin films of gold prepared by vacuum evaporation technique at very slow deposition rate contain various types of nanoparticles in it. The film with thickness < 3 nm has nearly spherical shape nanoparticles, whereas films with thickness ≥ 6 nm are dominated by non-cylindrical nanorods. These films provide two resonance peaks, where both the peaks follow an expression reported earlier and is found within the limit of Maxwell Garnet theory. However both the peaks fit for different refractive index of the surrounding medium. The LSPR peak wavelengths follow a linear relation with aspect ratio of the nanoparticles and best fit the expression derived by Link et al. [25] for only those nanoparticles having aspect ratio more than two. The LSPR peaks of these films shift to higher wavelength side once it is coated with Methylene Blue dye due to change in effective refractive index of the surrounding medium. The refractive index sensitivity of these films is found ranging from 250 - 305 nm/RIU for the film thicknesses ranging from 3 to 10 nm respectively. Sensitivity is saturating towards higher film thickness. These sensitivities are comparable to the sensitivities of gold nanorods (cylindrical shape) as well as nanoporous gold thin films. This show that film prepared using comparatively easy technique of vacuum evaporation can also be used as refractive index sensor for any organic or biological samples. However further investigation is needed to find different nanostructures as well as the techniques to increase the refractive index sensitivity of thin films in visible region.

**TABLE-1** Data of surface roughness measurement of gold thin film using AFM

| Gold Film Thickness (t) | Pk-Pk Roughness (nm) | Average Roughness (nm) | RMS Roughness (nm) |
|---|---|---|---|
| 3 nm (1x1 μm) | 3.05 | 0.26 | 0.34 |
| 6 nm (1x1 μm) | 6.10 | 0.51 | 0.69 |

**TABLE-2** Experimental data of aspect ratio and LSPR peak wavelengths along with calculated values of $\varepsilon_m$ using eq.-1

| S. No. | Aspect Ratio (R) | $\lambda_m$ (nm) | $\varepsilon_m$ |
|---|---|---|---|
| 1. | 0.42 | 526 | -1.66 |
| 2. | 0.82 | 582 | -5.78 |
| 3. | 1.54 | 601 | 25.57 |
| 4. | 2.14 | 650 | 7.09 |
| 5. | 3.52 | 710 | 3.51 |
| 6. | 5.88 | 790 | 2.12 |
| 7. | 8.33 | 962 | 2.12 |

**TABLE-3** Data of refractive index and refractive index sensitivity for different film thickness

| S. No. | Film thickness (nm) | Volume Fraction (F) | $\lambda_m$ (Au) (nm) | $n_s$ (Au) | $\lambda_m$(Au+MB) | $n_s$ (Au+MB) | $\Delta\lambda/\Delta n_s$ |
|---|---|---|---|---|---|---|---|
| 1 | 3 | 0.40 | 635 | 2.37 | 670 | 2.51 | 250 |
| 2 | 6 | 0.51 | 677 | 2.24 | 733 | 2.43 | 295 |
| 3 | 10 | 0.53 | 736 | 2.38 | 791 | 2.57 | 306 |

**Figure Caption**

1. Absorption spectra of the gold thin films of various thicknesses, (1) 0.5, (2) 1, (3) 2, (4) 3, (5) 3.4, (6) 6 and (7) 10 nm.

2. Variation in the localized surface plasmon resonance (LSPR) peak wavelengths with film thicknesses.

3. Absorption spectra of gold thin films and its composite film having Methylene Blue dye coating, (1) 3 nm gold, (2) 3nm gold + MB, (3) 6 nm gold, (4) 6 nm gold + MB, (5) 10 nm gold, (6) 10 nm gold + MB

4. Variation in LSPR peak wavelength with film thickness: (1) Gold film, (2) Composite film of gold and Methylene Blue

5. Variation in the size and the shape of nanoparticles in the gold film of 1 nm thickness, (a) TEM micrograph, (b) Histogram of variation in the ellipticity of the nanoparticles on substrate.

6. Variation in the size and the shape of nanoparticles in the gold film of 3 nm thickness, (a) TEM micrograph, (b) Histogram of variation in the ellipticity of the nanoparticles on substrate.

7. Variation in the size and the shape of nanoparticles in the gold film of 6 nm thickness, (a) TEM micrograph, (b) Histogram of variation in the ellipticity of the nanoparticles on substrate.

8. AFM micrograph of the surface structures in 3 nm thick gold film, (a) 2 D picture, (b) 3 D picture.

9. AFM micrograph of the surface structures in 6 nm thick gold film, (a) 2 D picture, (b) 3D picture.

10. Variation in LSPR peak wavelength with aspect ratio calculated using eq.-1:

    (1) $\varepsilon_m$=1, (2) $\varepsilon_m$=4, (3) $\varepsilon_m$=9, (4) $\varepsilon_m$=16

11. Variation in LSPR peak wavelength with volume fraction calculated using eq.-2.

    (1) $n_s$=1, (2) $n_s$=1.5, (3) $n_s$=2, (4) $n_s$=2.5, (5) $n_s$=3,

    Black triangle and squares are experimental data points for $\lambda_1$ and $\lambda_2$ peaks of gold thin films.

12. Variation in LSPR peak wavelength with volume fraction calculated using eq.-2. (2) $n_s=1$, (2) $n_s=1.5$, (3) $n_s=2$, (4) $n_s=2.5$, (5) $n_s=3$,

Black squares are experimental data points for $\lambda_2$ peaks of composite film of gold and MB.

13. Variation in refractive index sensitivity of thin films with film thickness

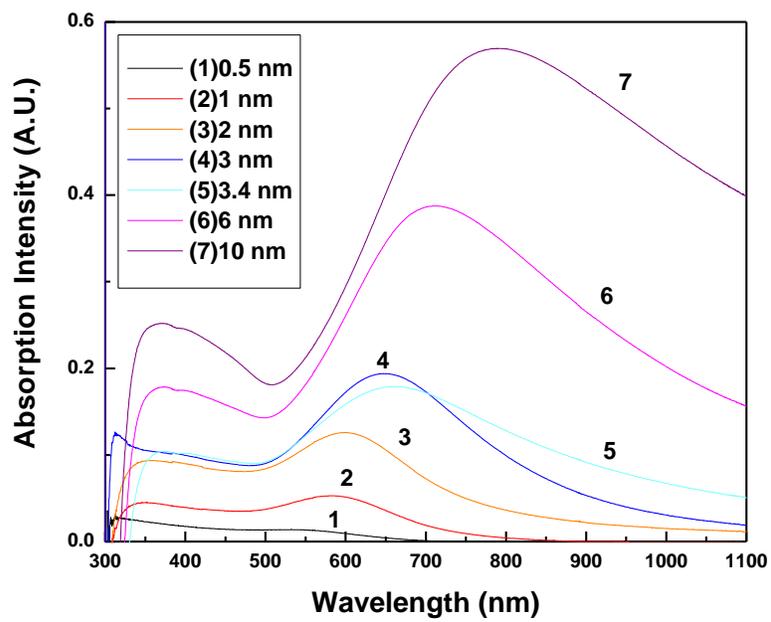

**Fig.-1**

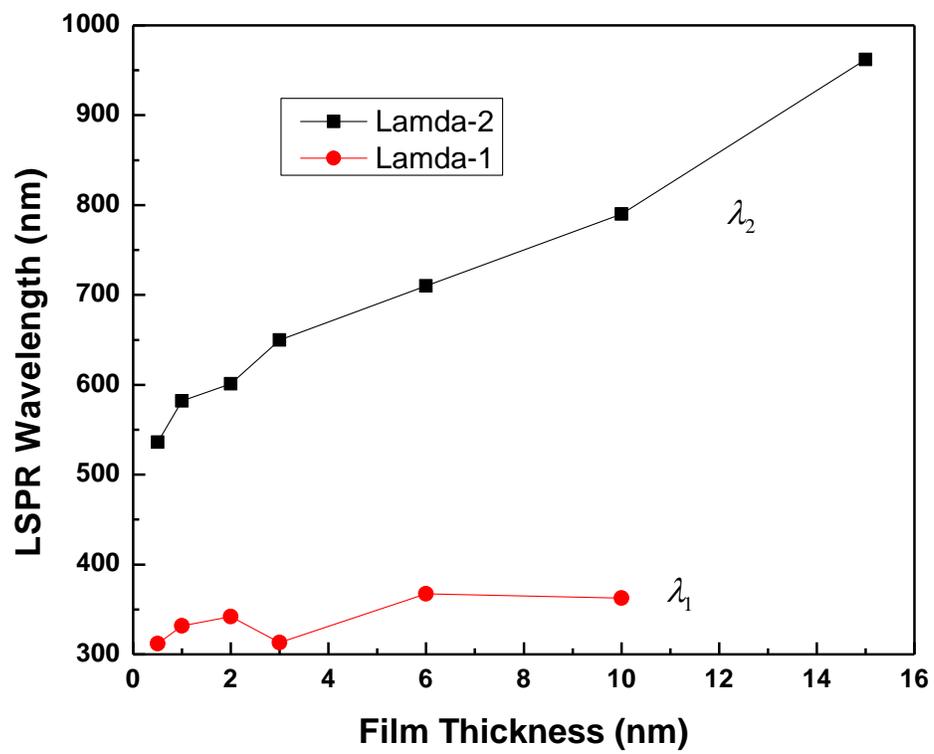

**Fig.-2**

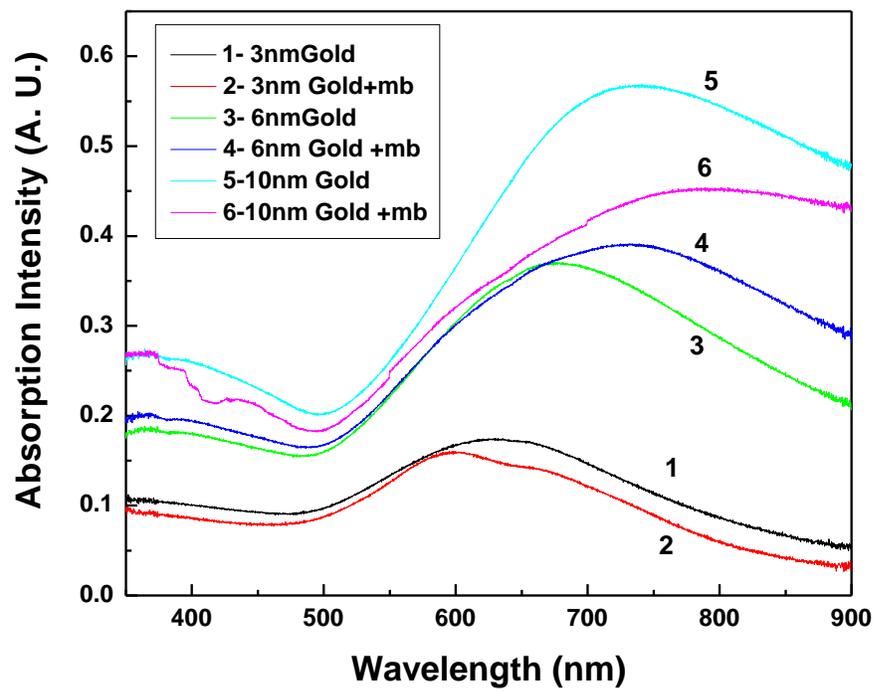

**Fig.- 3**

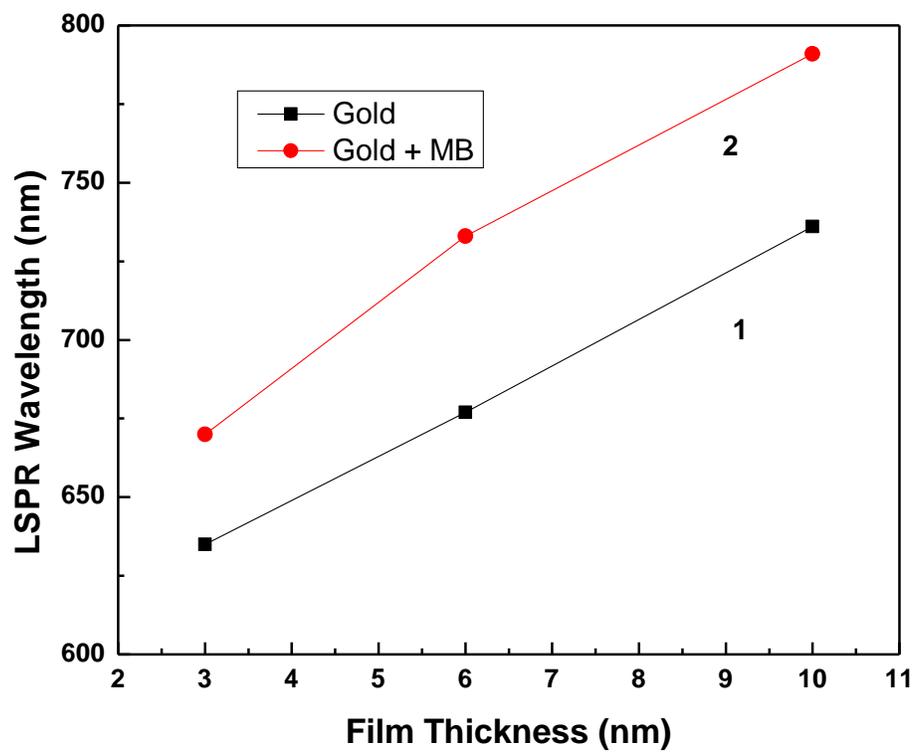

**Fig.-4**

**(a)**

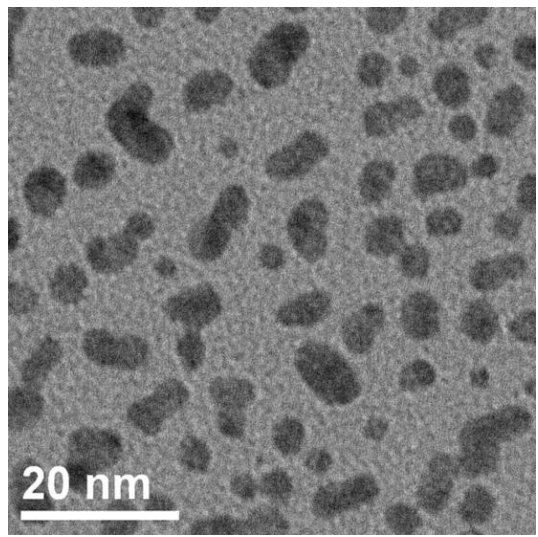

**(b)**

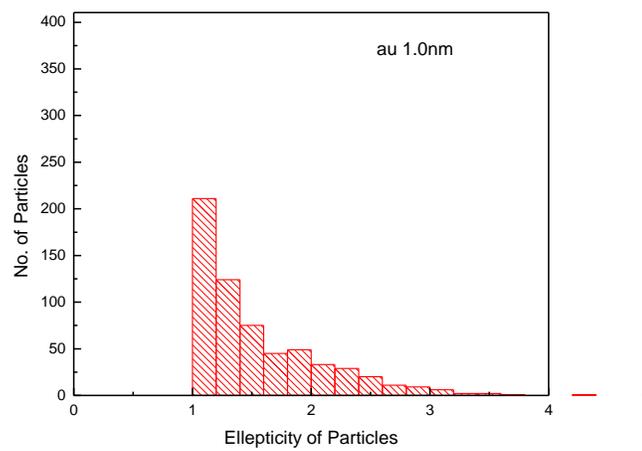

**Fig.-5**

**(a)**

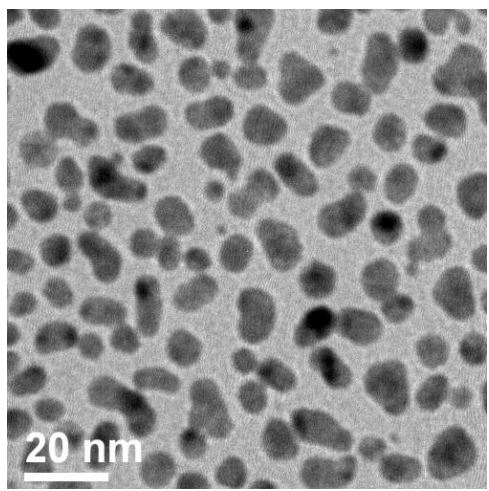

**(b)**

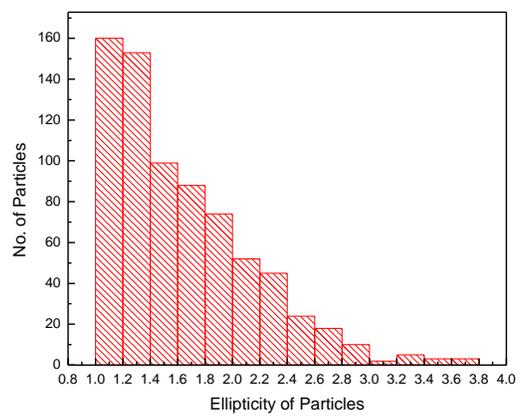

**Fig.-6**

**(a)**

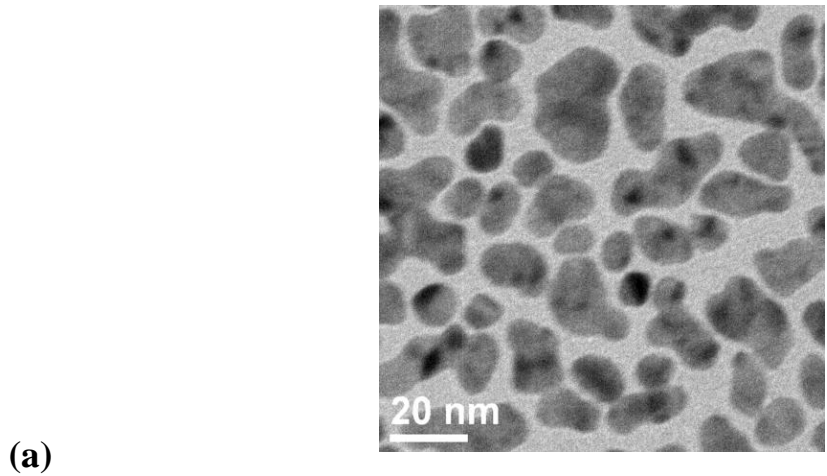

**(b)**

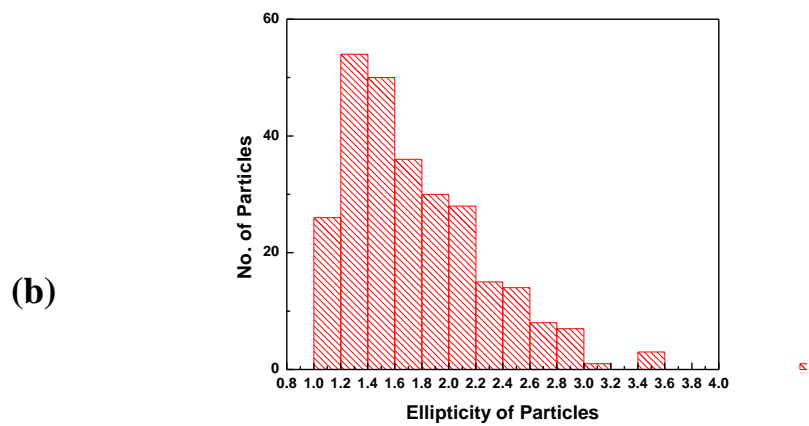

**Fig.-7**

**(a)**

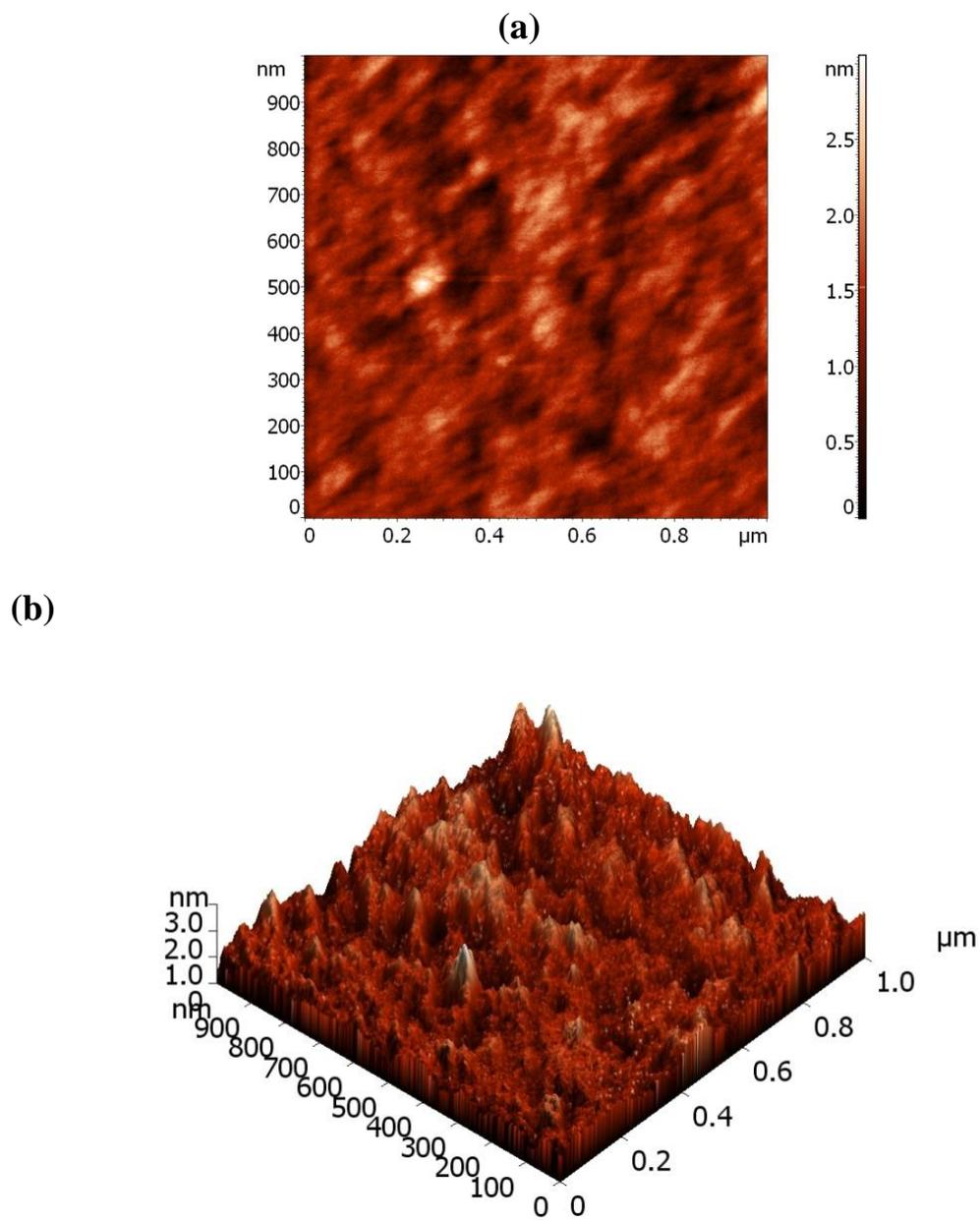

**(b)**

**Fig.-8**

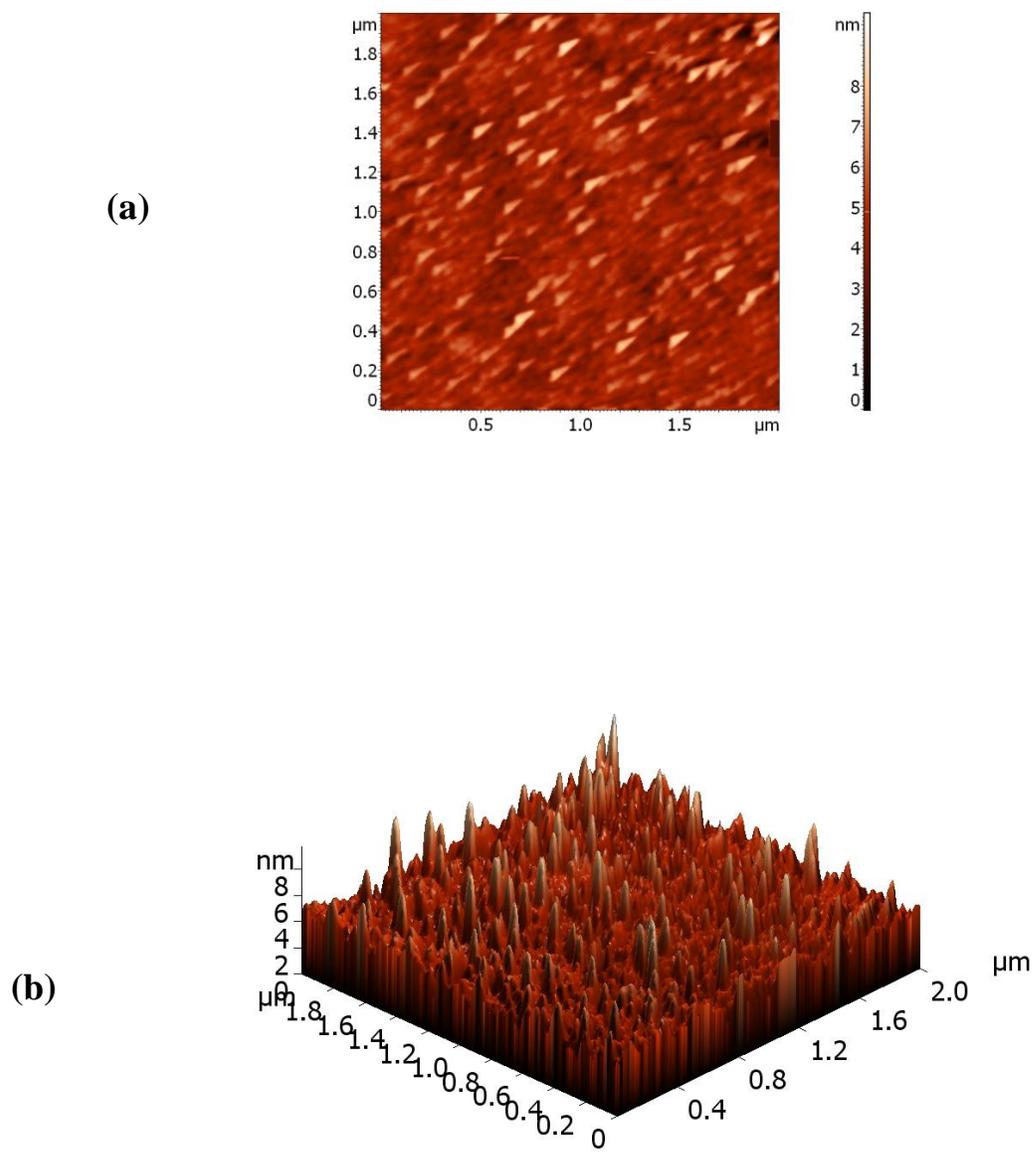

**(a)**

**(b)**

**Fig.-9**

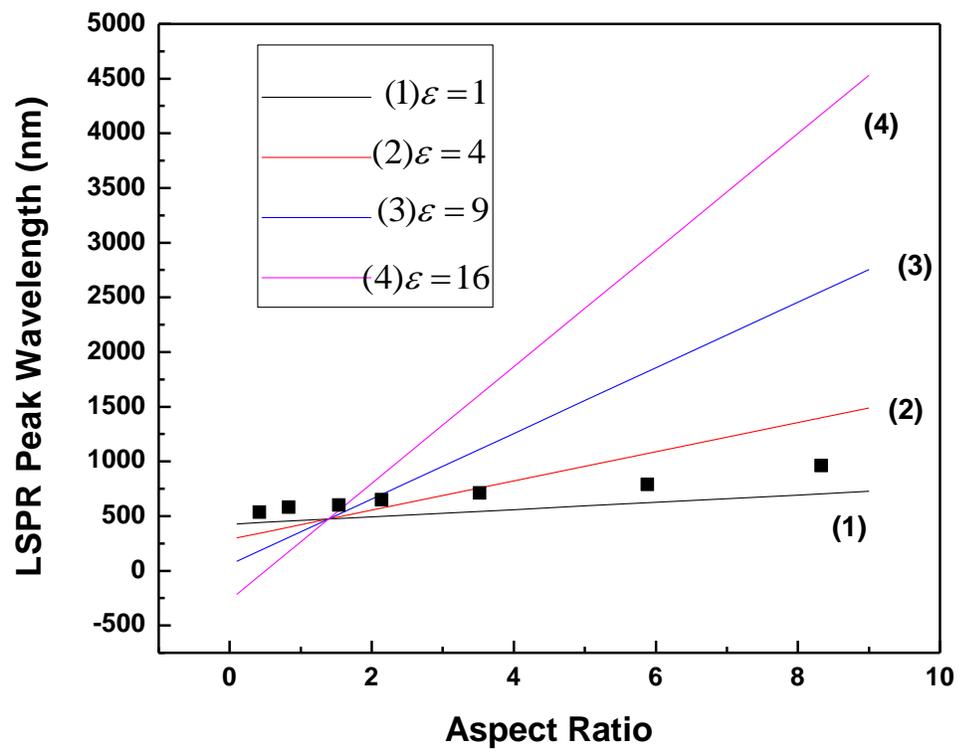

**Fig.-10**

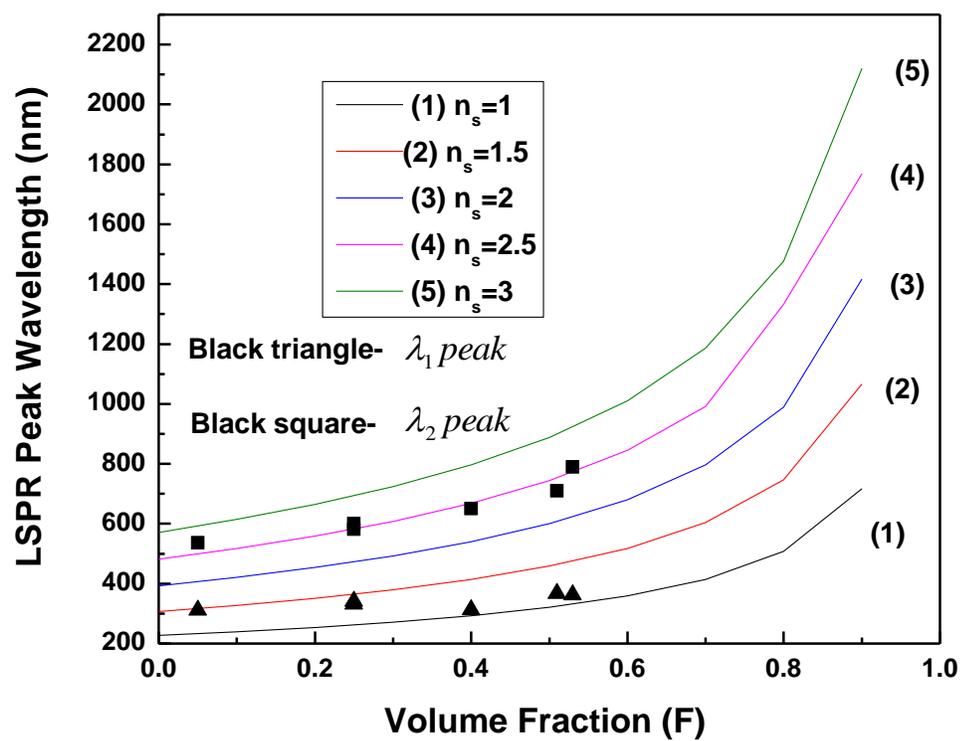

**Fig.-11**

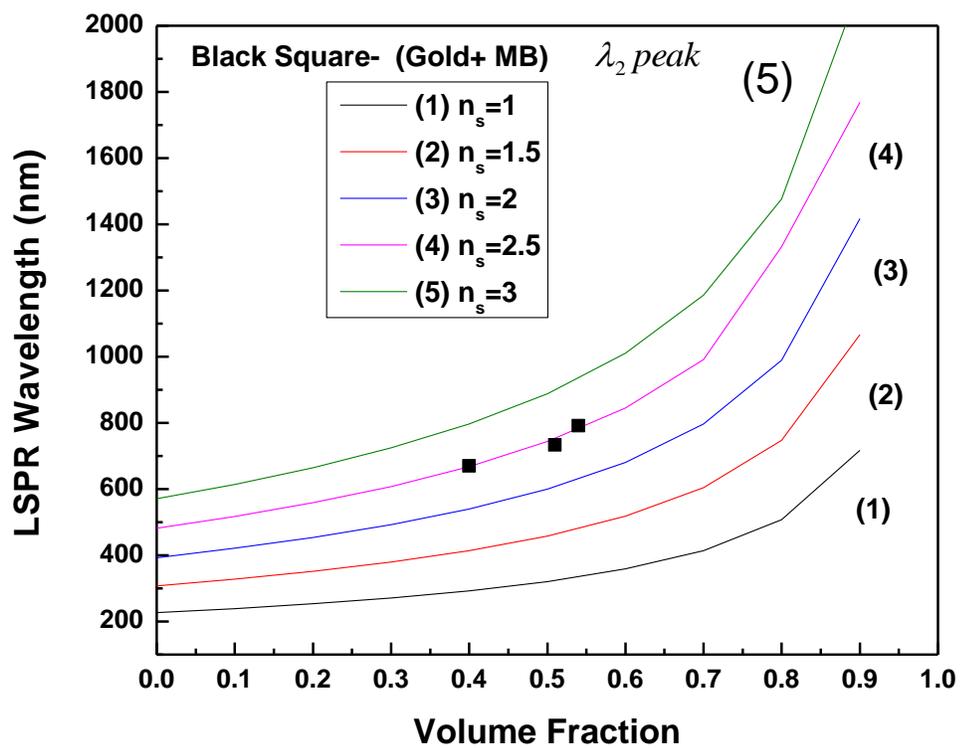

**Fig.-12**

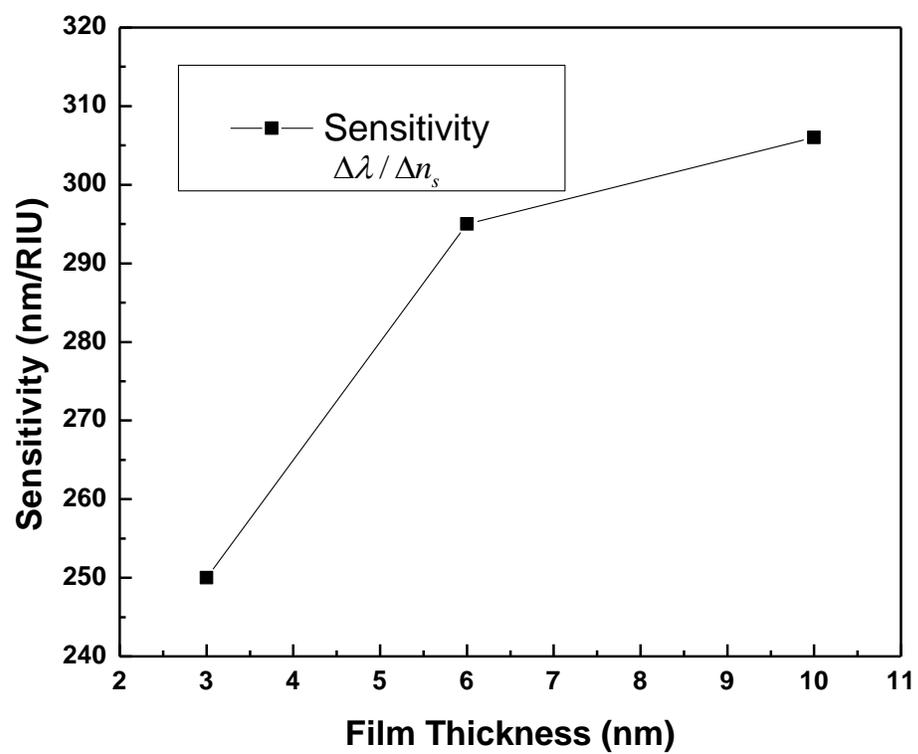

**Fig.-13**